\documentclass[conference, 10pt]{IEEEtran}
\IEEEoverridecommandlockouts 

\usepackage{amsmath}  
\usepackage{amssymb,amsfonts}
\usepackage{graphicx}

\usepackage{ifxetex}
\ifxetex
\else
\usepackage[utf8]{inputenc}
\usepackage[T1]{fontenc}
\usepackage{microtype}
\fi

\usepackage[boxed,vlined]{algorithm2e}

\newtheorem{theorem}{Theorem}
\newtheorem{remark}{Remark}
\newtheorem{lemma}{Lemma}

\newtheorem{proposition}{Proposition}

\def\gap{1.55ex}
  \abovedisplayskip\gap
  \belowdisplayskip\gap
  \abovedisplayshortskip\gap
  \belowdisplayshortskip\gap

\graphicspath{{../pictures/}}

\begin{document}

\title{Optimal Coding Functions for Pairwise Message Sharing on Finite-Field Multi-Way Relay Channels}
\author{\IEEEauthorblockN{Lawrence Ong, Sarah J.\ Johnson, Christopher M.\ Kellett}
\IEEEauthorblockA{School of Electrical Engineering and Computer Science, 
The University of Newcastle, Australia}
\thanks{Lawrence Ong is the recipient of an ARC Discovery Early Career Researcher Award (DE120100246). Sarah Johnson and Christopher Kellett are recipients
of ARC Future Fellowships (FT110100195 and FT110100746 respectively).}}
\maketitle

\begin{abstract}
This paper considers the finite-field multi-way relay channel with pairwise message sharing, where multiple users exchange messages through a single relay and where the users may share parts of their source messages (meaning that some message parts are known/common to more than one user). In this paper, we design an optimal functional-decode-forward coding scheme that takes the shared messages into account. More specifically, we design an optimal function for the relay to decode (from the users on the uplink) and forward (back to the users on the downlink).
We then show that this proposed function-decode-forward coding scheme can achieve the capacity region of the finite-field multi-way relay channel with pairwise message sharing.
This paper generalizes our previous result for the case of three users to any number of users.
\end{abstract}


\section{Introduction}

\begin{figure}
\centering
\includegraphics[height=2cm]{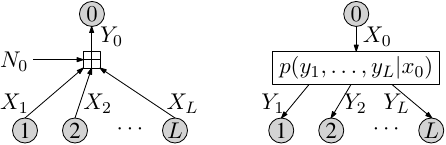}
\caption{The multi-way relay channel with finite-field uplink (left) and arbitrary downlink (right), where users (nodes 1, 2, $\dotsc$, $L$) exchange messages through a relay (node 0).}
\label{fig}
\end{figure}

In this paper, we consider the finite-field multi-way relay channel where multiple users exchange messages through a relay, and where the channel from the users to the relay is modeled by a finite-field channel (see Figure~\ref{fig}).

The finite-field multi-way relay channel with \emph{independent messages} was first considered in \cite{ongmjohnsonit11} where we proposed an optimal scheme that achieves the capacity. The scheme is, however, suboptimal when the messages are \emph{correlated}, e.g., when there is message \emph{sharing} (a special correlation structure in which every user pair may have part of their messages in common). See Appendix~\ref{section:example} for a proof of the suboptimality of the scheme
in \cite{ongmjohnsonit11} when there are shared messages.
The shortcoming of this scheme  is that it ignores the fact that some parts of the messages are known to two users (i.e., the shared parts).
 For the special case of three users with message sharing, an optimal function was constructed by Ong et al.~\cite{onglechnerjohnsonkellett13}. However, the function is specific to three users, and is not extendable to more users. Indeed, prior to the current paper, it was not clear if optimal functions existed for an arbitrary number of users.

In this paper, we design a scheme that constructs an optimal function for \emph{any number} of users with pairwise message sharing, thereby obtaining the capacity of the channel.

\subsection{A Challenge: To Design an Optimal Function to be Decoded by the Relay}

The multi-way relay channel models many relay-aided communication networks where there is no direct user-to-user link, e.g., cellular mobile networks and satellite communications.

Theoretically, the multi-way relay channel poses new challenges in multi-user information theory not encountered in the classical setups~\cite{elgamalkim2001}, e.g., the multi-access channel, the broadcast channel, the interference channel, and the relay channel. Although a relay is present in both the relay channel and the multi-way relay channel, the latter involves data transmission in multiple directions. 

Classical relaying schemes~\cite{kramergastpar04} (decode-forward, amplify-forward, compress-forward) can be used for the multi-way relay channel. However, the functional-decode-forward scheme (also known as compute-forward~\cite{nazergastpar11}) outperforms the classical relaying schemes in certain network configurations~\cite{gunduzgoldsmithpoor-13-it}, and even achieves the capacity when the channel is a finite-field channel~\cite{ongmjohnsonit11}.

The functional-decode-forward scheme incorporates network coding~\cite{ahlswedecai00}
in the design of the channel codes to facilitate bidirectional relaying. For example, consider the two-way relay channel, where node 1 sends its message bit $W_1$ to node 2 via the relay, and node 2 sends $W_2$ to node 1 via the relay. The relay decodes a function $f=W_1 \oplus W_2$ directly on the \emph{uplink} (i.e., the channel from the users to the relay), as opposed to decoding both message bits individually, and then performing the XOR operation $\oplus$. It then re-encodes and broadcasts $f$ on the \emph{downlink} (i.e., the channel from the relay to the users). 

Although the functional-decode-forward scheme is a good candidate for the multi-way relay channel, the challenge is to find an optimal function for the relay to decode. Three major difficulties are as follows: (i) The function should be \emph{matched} to the channel such that efficient channel codes can be used for the users while allowing the relay to decode the function. (ii) The function should contain the least information that the relay needs to decode, or equivalently, it should allow the users to transmit the most information. (iii) The function, when broadcast back to the users, must allow each user to decode its required messages.

To illustrate the first difficulty, consider the two-way relay channel with an additive white Gaussian noise (AWGN) uplink $Y_0 = X_1 + X_2 + N_0$, where $Y_0$ is the received signal at the relay, and $N_0$ is zero-mean additive white Gaussian noise with variance $\sigma^2$. Suppose that the users $a \in \{1,2\}$ each transmit an $nR$-bit message, $W_a$, in $n$ uplink channel uses. A good function is $f = L(W_1) + L(W_2) \mod \Lambda$, where $L(\cdot)$ is a one-to-one function that maps the message to a high-dimensional \emph{lattice point}, and $\!\!\!\mod \Lambda$ is a modulo operation defined with respect to the lattice~\cite{namchunglee09}. Nam et al.~\cite{namchunglee09} have shown that the relay can \emph{reliably} (i.e., with an arbitrarily small error probability as $n$ increases) decode the function~$f$ if $R < \frac{1}{2} \log_2 \left( \frac{1}{2} + \frac{P}{\sigma^2} \right)$, where $P$ is the average transmitted power of each user, and $R$ is the rate (message bits per channel use) at which each user transmits. A shortcoming of this scheme is that each user cannot transmit at its point-to-point channel (user-to-relay) capacity $\frac{1}{2} \log_2 \left( 1 + \frac{P}{\sigma^2}\right)$ (see the aforementioned challenge (i)). However, this scheme satisfies challenges (ii) and (iii), and it  matches the AWGN channel well in the high signal-to-noise ratio, $\frac{P}{\sigma^2}$, regime. The channel-matching difficulty is accentuated in the general uplink channel where it is not clear if efficient channel codes exist that allow each user to transmit at its user-to-relay channel capacity while simultaneously allowing the relay to decode the required function.

In this paper, we will focus on challenges (ii) and (iii). To this end, we consider the finite-field uplink. Using random linear block codes to match the channel, each user is able to transmit at its point-to-point channel capacity while the relay is able to decode the summation of all users' messages. Although the finite-field channel simplifies challenge (i), finding the optimal function is not trivial.

The finite-field model captures two important characteristics of the AWGN channel: 
Firstly, the channel is corrupted with additive noise; secondly, the channel inputs interfere linearly with each other, i.e., the transmitted signals are summed at the receiver. 
Sharing these two properties, optimal coding schemes derived for the finite-field channel shed light on how one would code in AWGN channels. For example, the optimal coding scheme derived for the finite-field multi-way relay cahnnel with independent sources~\cite{ongmjohnsonit11} has been used to prove capacity results for the AWGN counterpart~\cite{ongkellettjohnson12it}.

Sources with shared messages can model networks with different data types. For example, consider a sensor network where each sensor measures the temperature and the wind direction. The temperature measured by each sensor may be independent, but some sensor pairs may always record the same wind direction.

\section{Main Results}

\subsection{Channel Model}
Figure~\ref{fig} depicts the finite-field multi-way relay channel with users denoted by 1, 2, $\dotsc$, $L$, and the relay denoted by 0. The channel input from node $i$ is given by $X_i$, and the channel output received by node $i$ by $Y_i$. The channel consists of a finite-field uplink, defined as
\begin{equation}
Y_0 = X_1 \oplus X_2 \oplus \dotsm \oplus X_L \oplus N_0 \triangleq \bigoplus_{i=1}^L X_i \oplus N_0, \label{eq:uplink}
\end{equation}
and an \emph{arbitrary} downlink (not restricted to a finite field),
\begin{equation}
p_{Y_1,Y_2,\dotsc,Y_L|X_0}(y_1,y_2,\dotsc,y_L|x_0). \label{eq:downlink}
\end{equation}
The uplink variables $Y_0$, $N_0$, and $X_i$ for all $i \in \{1,2,\dotsc,L\}$ are elements of a finite field of any order $F$, $\oplus$ is addition in the field, and $N_0$ is arbitrarily-distributed noise. 
We consider the general downlink model, which includes the finite-field downlink~\cite{ongmjohnsonit11} as a special case.


We consider pairwise message sharing where there are two types of messages: private messages and shared messages. Each private message $W_a$ is known to only user $a$, and each shared message $W_{a,b}$ (where $a \neq b$) is known to two users $a$ and $b$, for $a,b \in \{1,2,\dotsc, L\}$. The terms ``private'' and ``shared'' here are defined with respect to the senders. So, there are all together $L + \binom{L}{2}$ independent messages $W_\mathcal{I}$, for $\mathcal{I} \in \Big\{$all singletons $\{1\},\{2\},\dotsc, \{L\},$  and all unordered pairs $\{a,b\}$ where $a,b \in \{1,2,\dotsc,L\}$ and $a \neq b \Big\}$. This also means each user $a$ has/knows one private message $W_a$ and $(L-1)$ shared messages $\{W_{a,i}: i \in \{1,2,\dotsc,L\} \setminus \{a\}\}$ (each shared with one other user $i$). Note that $W_{a,b} = W_{b,a}$ by definition.

We consider full message exchange where each user $a$ needs to decode all $\{W_\mathcal{I}: a \notin \mathcal{I}\}$, i.e., all messages unknown to $a$. We use the channel $n$ times for the users to exchange these messages.
Let each $W_\mathcal{I}$ be an $nR_\mathcal{I}$-bit message, where $R_\mathcal{I}$ (bits per channel use) is the rate of message $W_\mathcal{I}$ transmitted.
With this, we define the rate tuple $\boldsymbol{R} \triangleq ( R_1, R_2, \dotsc, R_L, R_{1,2}, R_{1,3}, \dotsc, R_{1,L}, R_{2,3}, R_{2,4}, \dotsc, R_{2,L},$ $\dotsc, R_{L-1,L})$, i.e., the collection of all message rates.


\subsection{Results}

We define a length-$n$ block code as follows, where we use square brackets to indicate the variable associated with a channel use:  
\begin{itemize}
\item Encoding for each user $a$: $X_a[t] = f_{at} \big( W_a, (W_{a,i})_{i \in \{1,2\dotsc, L\}\setminus{a}}, (Y_a[\tau])_{\tau \in \{1,2,\dotsc, t-1\}} \big)$, for all $t \in \{1,\dotsc,n\}$, meaning that each user transmits its signal based on its known messages and previously received signals.
\item Encoding for the relay: $X_0[t] = f_{0t} \big( (Y_0[\tau])_{\tau \in \{1,2,\dotsc, t-1\}} \big)$, for all $t \in \{1,\dotsc,n\}$,  meaning that the relay transmits each signal based on its previously received signals.
\item Decoding for each user $a$: $\big( (\hat{W}_{i(a)})_{i \in \{1,2,\dotsc, L\} \setminus \{a\}}, (\hat{W}_{\{i,j\}(a)})_{i,j \in \{1,2,\dotsc, L\} \setminus \{a\}} \big) = g_a \big( W_a, (W_{a,i})_{i \in \{1,2\dotsc, L\}\setminus{a}},Y_a[\tau])_{\tau \in \{ 1,2, \dotsc, n\}} \big)$, meaning that each user decodes its required messages based on all $n$ received signals and its known messages.
\end{itemize}
Here, $\hat{W}_{\mathcal{I}(a)}$ is the estimate of $\hat{W}_\mathcal{I}$ by user $a$. A decoding error occurs if some user wrongly decodes some messages. 
We denote the probability of decoding error by $P_\text{e}$. We say that the rate tuple $\boldsymbol{R}$ is achievable if the following is true: for any $\epsilon >0$, there exists a block code such that $P_\text{e} \leq \epsilon$. The capacity region is the closure of all achievable rate tuples.

To simplify notation, we define the sum rate of the  messages that node $a$ needs to decode as
\begin{equation}
R^\Sigma_a \triangleq \sum_{i \in \{1,2,\dotsc, L\} \setminus \{a\}} R_i + \sum_{ \{i,j\} \subset \{1,2,\dotsc, L\} \setminus \{a\}} R_{i,j}, \label{eq:sum-notation}
\end{equation}
and the largest sum rate to be decoded (by some user) as
\begin{equation}
R^\Sigma_\text{max} \triangleq \min_{i \in \{1,2,\dotsc, L\}} R^\Sigma_i. \label{eq:max-sum-rate}
\end{equation}

We now present the main result of this paper:
\begin{theorem}\label{theorem:main}
The rate tuple $\boldsymbol{R}$ is achievable if
\begin{equation}
R^\Sigma_\text{max} < \log_2 F - H(N_0), \label{eq:common-capacity-1}
\end{equation}
and there exists some $p(x_0)$ such that
\begin{equation}
R^\Sigma_a  < I(X_0;Y_a), \quad\quad \text{for all } a \in \{1,2,\dotsc, L\}. \label{eq:common-capacity-2}
\end{equation}
Conversely, if $\boldsymbol{R}$ is achievable, then there exists some $p(x_0)$ such that \eqref{eq:common-capacity-1} and \eqref{eq:common-capacity-2} hold with non-strict inequalities  (i.e., $\leq$ instead of $<$).
\end{theorem}

We will first propose a coding scheme for this channel in Section~\ref{sec:optimal-coding}. In Section~\ref{sec:proof}, we  will then show that our proposed coding scheme is optimal, thus giving us the capacity result in Theorem~\ref{theorem:main}.

\begin{table*}[t]
\renewcommand{\arraystretch}{1.3}
\caption{Uplink Message Transmission}
\label{table:uplink}
\begin{center}
\begin{tabular}{c|c|c|c|c|c|c|c|c|c|c|c|c|c|c|}
\multicolumn{1}{c}{ block } & \multicolumn{1}{c}{$2$} & \multicolumn{1}{c}{$3$} & \multicolumn{1}{c}{$\dotsm$} & \multicolumn{1}{c}{$L$} & \multicolumn{1}{c}{$(2,3)$} &  \multicolumn{1}{c}{ $(2,4)$} & \multicolumn{1}{c}{$\dotsm$} & \multicolumn{1}{c}{$(2,L)$} & \multicolumn{1}{c}{$(3,3)$} & \multicolumn{1}{c}{$(3,4)$} & \multicolumn{1}{c}{$\dotsm$} & \multicolumn{1}{c}{$(3,L)$} & \multicolumn{1}{c}{$\dotsm$} & \multicolumn{1}{c}{ $(L-1,L)$} \\
\multicolumn{1}{c}{ block size} & \multicolumn{1}{c}{$\ell_2$} & \multicolumn{1}{c}{$\ell_3$} & \multicolumn{1}{c}{$\dotsm$} & \multicolumn{1}{c}{$\ell_L$} & \multicolumn{1}{c}{$\ell_{2,3}$} &  \multicolumn{1}{c}{ $\ell_{2,4}$} & \multicolumn{1}{c}{$\dotsm$} & \multicolumn{1}{c}{$\ell_{2,L}$} & \multicolumn{1}{c}{$\ell_{3,3}$} & \multicolumn{1}{c}{$\ell_{3,4}$} & \multicolumn{1}{c}{$\dotsm$} & \multicolumn{1}{c}{$\ell_{3,L}$} & \multicolumn{1}{c}{$\dotsm$} & \multicolumn{1}{c}{ $\ell_{L-1,L}$} \\
\cline{2-15}
row 1 & $W_2$ & $W_3$ & $\dotsm$ & $W_L$ & $W_{2,3}$ &   $W_{2,4}$ & $\dotsm$ & $W_{2,L}$ & $W_{3,3}$ & $W_{3,4}$ & $\dotsm$ &$W_{3,L}$ & $\dotsm$ & $W_{L-1,L}$ \\
\cline{2-15}
row 2 & $*$ & & & & $*$ & $*$ & $\dotsm$ & $*$ & & & & & & \\
\cline{2-15}
row 3 & & $*$ & & & $*$ & & & & $*$ & $*$ & $\dotsm$ & $*$ & & \\
\cline{2-15}
$\vdots$ & \multicolumn{4}{l}{}  & \multicolumn{10}{l|}{}\\
\cline{2-15}
row $L$ & & & & $*$ & & & & $*$ & & & & $*$  & &  $*$ \\
\cline{2-15}
\end{tabular}
\end{center}

{\raggedleft Note: The cells are drawn as rectangles in the table. Each block contains $L$ cells (from row 1 to row $L$). Each cell contains multiple columns (not drawn in the table). Each column in the cell carries one message symbol.}
\end{table*}

\section{The Construction of Optimal Coding Functions for Pairwise Message Sharing} \label{sec:optimal-coding}

In this section, we will propose a coding scheme based on random linear block codes. To encode using linear block codes, we first convert each message $W_\mathcal{I}$ of $nR_\mathcal{I}$-bits to a finite-field vector of length $\ell_\mathcal{I}$, where $F^{\ell_\mathcal{I}} = 2^{nR_\mathcal{I}}$, or equivalently, $R_\mathcal{I} = \frac{\ell_\mathcal{I} \log_2 F}{n}$ and  $\ell_\mathcal{I} = \frac{nR_\mathcal{I}}{\log_2 F}$. 

Using the same notation as in \eqref{eq:sum-notation}, we define $\ell^\Sigma_a \triangleq \frac{nR^\Sigma_a}{\log_2 F}$ as the total number of finite-field symbols user $a$ needs to decode. 

Without loss of generality, suppose $\ell^\Sigma_1 = \frac{nR^\Sigma_\text{max}}{\log_2 F}$, meaning that
\begin{equation}
\ell^\Sigma_1 \geq \ell^\Sigma_a,\quad \text{for all } a \in \{2,3,\dotsc, L\}, \label{eq:simplify}
\end{equation}
i.e., among all users, user 1 is to decode the largest number of finite-field symbols.

\subsection{Constructing the Messages to be Transmitted by the Users and the Function $\boldsymbol{f}$ to be Decoded by the Relay}



The messages to be transmitted by the users (on the uplink) are described using Table~\ref{table:uplink}. The table consists of $L(L-1)/2$ blocks and $L$ rows, giving $L^2(L-1)/2$ cells. Each cell is drawn as a rectangle in the table. For each block, the users are to simultaneously transmit the messages assigned to the cells. Each cell contains multiple columns; each column in the cell corresponds to a message symbol (columns are not drawn in the table). We refer to the number of columns in each cell as the block size, which is set to be the length of the message in the cell in row 1.



We construct Table~\ref{table:uplink} as follows: In row 1, we place all the messages that user 1 requires, i.e., $\big\{W_\mathcal{I}: 1 \notin \mathcal{I} \big\}$. 
We identify each block by the subscript $\mathcal{I}$ of the corresponding message $W_\mathcal{I}$ in row 1; the block size is thus $\ell_\mathcal{I}$. This gives the total size of all the blocks to be $\ell^\Sigma_1$; we will see later that this is the length of the function the relay decodes.

We now fill in the rest of the rows.
Of the $L(L-1)/2$ messages in row 1, user $a \in \{2,3,\dotsc, L\}$ knows $(L-1)$ of them a priori, namely, $\big\{W_a, W_{a,j}: j \in \{2,3,\dotsc, L\} \setminus \{a\} \big\}$ (i.e., messages with subscript ``$a$''). For each of these messages, we put an asterisk in the cell in the same block in row $a$.

Over the asterisked cells in row $a$, we will assign messages that user $a$ requires and user 1 knows. This means we will assign $\big\{W_1, W_{1,j}: j \in \{2,3,\dotsc, L\} \setminus \{a\} \big\}$ to these cells (replacing each subscript ``$a$'' of the messages in row 1 by ``1''). We will see later that this choice allows both users 1 and $a$ obtain their respective required messages from the sum. For illustration, we extract only these $(L-1)$ blocks for rows 1 and $a$, as shown in the following table. These are the blocks where we have placed asterisks in row $a$. Note that these extracted blocks may not be consecutive in Table~\ref{table:uplink}.\\

\begin{scriptsize}
\renewcommand{\arraystretch}{1.3}
\begin{tabular}{c|c|c|c|c|c|c|c|}
\cline{2-8}
row 1 & $W_a$ & $W_{a,2}$ & $\dotsm$ &  $W_{a,a-1}$ &  $W_{a,a+1}$  & $\dotsm$ & $W_{a,L}$ \\
\cline{2-8}
row $a$ & \multicolumn{7}{|l|}{ $W_1, W_{1,2}, \dotsc, W_{1,a-1}, W_{1,a+1}, \dotsc,  W_{1,L}$} \\
\cline{2-8}
\end{tabular}
\end{scriptsize}
\\

\noindent We spread the messages $\big\{W_1, W_{1,j}: j \in \{2,3,\dotsc, L\} \setminus \{a\} \big\}$  across the asterisked cells in row $a$. This is necessary because the messages have different lengths, and they may not align at the block level. Expanding \eqref{eq:simplify} using \eqref{eq:sum-notation} we have
\begin{equation}
\ell_a + \sum_{j \in \{2,3,\dotsc, L\} \setminus \{a\}} \ell_{a,j} \geq \ell_1 +  \sum_{j \in \{2,3,\dotsc, L\} \setminus \{a\}}  \ell_{1,j}, \label{eq:more-symbols}
\end{equation}
meaning that the $(L-1)$ messages can always fit into the corresponding $(L-1)$ asterisked cells on row $a$.
If the above equality is strict, the excess columns in the asterisked cell(s) will be left empty.

We repeat this for all $a \in \{2,3,\dotsc, L\}$. Doing this, in every block $\mathcal{I}$, only rows $a \in \mathcal{I}$ have asterisked cells. 


For each column, we sum (using finite-field addition) all the unique symbols across all rows. In other words, if the same symbol appears multiple times in a column, it will only be summed once---meaning that only one copy needs to be transmitted on the uplink. We then define $f_\mathcal{I}$ as the summation in block $\mathcal{I}$. The function that the relay decodes is denoted by $\boldsymbol{f} \triangleq \big( f_\mathcal{I}: 1 \notin \mathcal{I} \big)$, i.e., each symbol in the function is the summation of one column in Table~\ref{table:uplink}. Clearly, $\boldsymbol{f}$ has $\ell^\Sigma_1$ symbols.

\begin{remark}
Recall that one challenge to design a good function is that it should carry the minimum number of bits the relay needs to decode. Here, user 1 is required to decode the largest number of message symbols, i.e., $\ell^\Sigma_1$ symbols, through the relay, and therefore the relay must decode and forward some function containing at least $\ell^\Sigma_1$ symbols. So, the  key to solving this problem lies in finding this function $\boldsymbol{f}$ which (i) has $\ell^\Sigma_1$ symbols, (ii) can be decoded by the relay, and (iii) allows each user to decode its intended messages. We will show that our proposed $\boldsymbol{f}$ satisfies these conditions.
\end{remark}

\subsection{Some Properties of Table~\ref{table:uplink}} \label{section:properties}

We now prove a few properties of Table~\ref{table:uplink}.

\begin{proposition} \label{prop:user1}
User 1 knows the messages in rows 2 to $L$ a priori.
\end{proposition}

\begin{IEEEproof}
In each row $a$, we only assign either $W_1$ or $W_{1,i}$ in the asterisked cells.
\end{IEEEproof}

\begin{proposition} \label{prop:user-a}
Once user $a$, for some $a \in \{2,3,\dotsc, L\}$, knows the messages in row $a$, it also knows the messages in the asterisked cells in all other rows.
\end{proposition}

\begin{IEEEproof}
Suppose that user $a$ has decoded the messages in row $a$, i.e., $\{W_1, W_{1,i}: i \in \{2,3,\dotsc, L\} \setminus \{a\}\}$. By definition, it knows $W_{1,a}$ a priori. Since, any message in other rows $\{2,3,\dotsc, L\} \setminus \{a\}$ must be either $W_1$ or $W_{1,i}$, user $a$ knows those messages.
\end{IEEEproof}

\begin{proposition} \label{prop:all}
For any $a \in \{1, 2,3,\dotsc, L\}$, after decoding all messages in rows $a$ and 1, user $a$ will have decoded all the messages it requires.
\end{proposition}

\begin{IEEEproof}
Recall that user $a$ needs to decode all messages $\big\{W_\mathcal{I}: a \notin \mathcal{I} \big\}$. Clearly, row 1 contains all messages that user 1 needs to decode. Now, we prove the proposition for $2 \leq a \leq L$. For $|\mathcal{I}|=1$, $W_1$ appears in row $a$, and $\{W_2, W_3, \dotsc, W_L\}$ appear in row 1. For $|\mathcal{I}|=2$, all messages $\big\{W_{i,j}: i,j \in \{2,3,\dotsc, L\} \setminus \{a\}\big\} \triangleq \mathcal{W}'$ appear in row 1, and $\big\{W_{1,k}: k \in \{2, 3,\dotsc, L\} \setminus \{a\} \big\} \triangleq \mathcal{W}''$ in row $a$. So, $\mathcal{W}' \cup \mathcal{W}'' = \{W_{i,j}: i,j \in \{1,2,\dotsc, L\} \setminus \{a\}\}$.
\end{IEEEproof}



\subsection{Each User Can Decode Its Required Messages from the Function $\boldsymbol{f}$} \label{sec:decode_from_f}

Knowing all the messages in the asterisked cells a priori (Proposition~\ref{prop:user1}), user 1 can decode all messages in row 1 from $\boldsymbol{f}$, and hence obtain all its intended messages (Proposition~\ref{prop:all}). For other users, suppose that user $a \in \{2,3,\dotsc, L\}$ has decoded $\boldsymbol{f}$, it first attempts to decode the messages in the asterisked cells in row $a$. From Proposition~\ref{prop:user-a}, it would have decoded the messages in all other asterisked cells. This allows the user to decode the messages in row 1, and hence all its required messages (Proposition~\ref{prop:all}).

So, we only need to show that any user $a$ can decode the messages (in the asterisked cells) in row $a$, denoted by $W_\Lambda \triangleq \big\{W_1, W_{1,i}: i \in \{2,3,\dotsc, L\} \setminus \{a\} \big\}$. It does so from the relevant blocks of $\boldsymbol{f}$, namely, $\{f_\mathcal{I}: \mathcal{I} \in \Theta\}$, where $\Theta \triangleq \{a, (a,i): i \in \{2,3,\dotsc, L\} \setminus \{a\} \}$ are all the blocks with asterisked cells in row $a$. Recall that the messages $W_\Lambda$ may not align and can be shorter than the blocks $\Theta$. The blocks in $\Theta$ are re-drawn here:\\

\hspace{-5ex} \begin{scriptsize}
\renewcommand{\arraystretch}{1.3}
\begin{tabular}{c|c|c|c|c|c|c c c}
\cline{2-6}
row 1 & $W_a$ & $W_{a,2}$  & $W_{a,3}$  & $\dotsm$ & $W_{a,L}$ &  $\Big] W_\Theta$  & \\
\cline{2-6}\cline{8-8}
row 2 & & $*$ & & $\dotsm$ & & \multicolumn{2}{l|}{}  \\
\cline{2-6}
$\vdots$ & \multicolumn{5}{l|}{}  & \multicolumn{2}{l|}{}\\
\cline{2-6}
row $a$ & \multicolumn{5}{l|}{ $W_1, W_{1,2}, \dotsc, W_{1,L}\quad \textnormal{(all $*$ cells in row $a$})$ } & $\Big] W_\Lambda$  & \multicolumn{1}{l|}{}\\
\cline{2-6}
$\vdots$ & \multicolumn{5}{l|}{}  & \multicolumn{2}{l|}{} & $V_\Theta$\\
\cline{2-6}
row $L$ & & & &  $\dotsm$ & $*$ & \multicolumn{2}{l|}{}\\
\cline{2-6} \cline{8-8}
\end{tabular}
\end{scriptsize}\\

For each block $\mathcal{I} \in \Theta$, user $a$ knows $W_\mathcal{I}$ (in row 1). It subtracts these from $f_\mathcal{I}$ to obtain $V_\mathcal{I} \triangleq f_\mathcal{I} - W_\mathcal{I}$, where $V_\mathcal{I}$ is the summation of all messages from rows 2 to $L$ (recall that only unique messages are added).

 Note that $W_\Lambda$ comprises $ \ell_1  + \sum_{i \in \{2,3,\dotsc, L\} \setminus \{a\} }\ell_{1,i} \triangleq A'$ symbols. Denote the first $A'$ symbols of $V_\Theta$ by $V_\Theta'$, which are functions of only $W_\Lambda$ (Proposition~\ref{prop:user-a}). 

So, user $a$ can decode $W_\Lambda$ from  $V_\Theta'$ if $V_\Theta'$  forms $A'$ linearly independent equations. We now propose an algorithm to rearrange the messages in $\boldsymbol{f}$ to achieve this.

\subsection{An Algorithm to Shuffle the Messages} \label{sec:algorithm}

Our aim is to get $A' \triangleq  \ell_1  + \sum_{i \in \{2,3,\dotsc, L\} \setminus \{a\} }\ell_{1,i}$ linearly independent equations in $V_\Theta'$. This must be true \emph{simultaneously} for all users $a \in \{2,3,\dotsc, L\}$. So, we ignore row 1 and consider asterisked cells in Table~\ref{table:uplink}. Each block has at most two asterisked cells. For each column from row 2 to row $L$, we denote the \emph{simplified column} by ${\alpha \brack \beta}$, where $\alpha$ and $\beta$ are symbols from the asterisked cells. If there is only one asterisked cell for that column, we have ${\alpha \brack }$.
Using this notation, we now propose the shuffling algorithm.
\begin{algorithm}[h]
\begin{small}
\Repeat{The \textbf{while} condition is not satisfied for one cycle of the \textbf{foreach} loop}{
\ForEach{user $a= 2, 3, \dotsc, L$} {
Consider all the non-empty columns in row $a$\; 
Rearrange each simplified column such that the top symbol takes the symbol in row $a$\;
\While{there exists two columns ${\alpha \brack \beta}$ and ${\gamma \brack \alpha}$ for some $\beta \neq \alpha$ and $\gamma \neq \alpha$}{
Swap $\alpha$ and $\gamma$ to get ${\gamma \brack \beta}$ and ${\alpha \brack \alpha}$\;
}
}
}
\caption{The Shuffling Algorithm}
\label{algo}
\end{small}
\end{algorithm}

Although swapping the symbols for one user may affect the simplified columns for other users, in each swap, we always increase the number of simplified columns with two identical symbols, i.e., ${\alpha \brack \alpha}$. As there are only a finite number of columns, the algorithm will always terminate after at most $\ell^\Sigma_1$ cycles of the \textbf{foreach} loop.

We will now show that when the algorithm ends, user $a$ can decode the messages on row $a$, i.e., $W_\Lambda$.  We treat ${\alpha \brack \alpha}$ as ${\alpha \brack }$ because only one copy of the same symbol is added to get $V_\Theta'$.
When we swap the symbols for each user in the algorithm, we always swap the top symbols, i.e., symbols in the same row---the bottom symbols may be from different rows. Hence, the properties of $\boldsymbol{f}$ derived in Section~\ref{section:properties} remain true.

Consider the decoding of user $a$ using the $A'$ non-empty simplified columns. The user again rearranges each simplified column such that the top symbol is from row $a$. Doing that, the top symbols of these simplified columns are distinct---they corresponds to the $A'$ symbols in $W_\Lambda$. Now,  if a symbol $\alpha$ appear at the top of a simplified column and the bottom of another, it has to take the form ${\alpha \brack }, {\beta \brack \alpha}, {\gamma \brack \alpha}, \dotsc, {\zeta \brack \alpha}$, because all $\alpha$ can only appear once at the top, and the case ${\alpha \brack \beta}$ and ${\gamma \brack \alpha}$ for any $\beta \neq \alpha$ and $\gamma \neq \alpha$ has been eliminated. Consequently, user $a$ can decode all the top symbols, i.e.,  $W_\Lambda$. This is true for all $a \in \{2,3,\dotsc, L\}$. So, we have shown that if a user can decode $\boldsymbol{f}$ (with message shuffling), then it can decode all its required messages.

\section{Proof of Theorem~\ref{theorem:main}} \label{sec:proof}

\subsection{The Converse}

The converse follows from the cut-set arguments~\cite[Thm 15.10.1]{coverthomas06}, \cite[eqs (11)--(12)]{mohajertiandiggavi10}, \cite[Sec III]{ongmjohnsonit11} and is omitted here. 

\subsection{Achievability}

\emph{Sketch of proof:} We pick any rate tuple $\boldsymbol{R}$ that satisfies \eqref{eq:common-capacity-1} and \eqref{eq:common-capacity-2}. To show that this rate tuple is achievable, we employ the length-$\ell^\Sigma_1$ finite-field vector $\boldsymbol{f}$, as defined above in Table~\ref{table:uplink} and shuffled with Algorithm~\ref{algo}, and show that
\begin{enumerate}
\item the relay is able to decode $\boldsymbol{f}$ in $n$ uplink uses,
\item each user is able to decode $\boldsymbol{f}$ broadcast by the relay in $n$ downlink uses, and
\item each user can decode its required messages from $\boldsymbol{f}$ and its own messages.
\end{enumerate}
Part 3) has been shown in Sections~\ref{sec:decode_from_f} and \ref{sec:algorithm}, and we will show parts 1) and 2) below.


\begin{remark}
In our proposed  coding scheme, the relay transmits after decoding $\boldsymbol{f}$. This means a total of $2n$ channel uses ($n$ for the uplink, followed by $n$ for the downlink). This issue can be easily rectified by repeating this scheme multiple times for multiple messages. Using the uplink and the downlink simultaneously, the relay transmits $\boldsymbol{f}$ that it has previously decoded, while, at the same time, the users transmit new messages.
This is a commonly-used technique for relay channels (see, e.g., \cite{covergamal79}\cite{ongmjohnsonit11}).
\end{remark}

\subsection*{1) The relay is able to decode $\boldsymbol{f}$ in $n$ uplink uses}

First, we quote  results for random linear block codes~\cite{onglechnerjohnsonkellett13}. 
\begin{lemma} \label{lemma:uplink}
Consider the channel \eqref{eq:uplink}. Suppose that each user $a \in \{1,2,\dotsc, L\}$ encodes a length-$\ell$ finite-field message (say $U_a$) into a length-$n$ finite-field codeword, and all users transmit their codewords simultaneously in $n$ channel uses. If $n$ is sufficiently large and if  $\frac{\ell \log_2 F}{n} < \log_2 F - H(N_0)$, then the relay can reliably decode $U_1 \oplus U_2 \oplus \dotsm \oplus U_L$. 
\end{lemma}

The transmission described in Table~\ref{table:uplink} utilizes $n$ uplink channel uses in total, and this is divided proportionally to all the blocks, i.e., each block $\mathcal{I}$ utilizes $n \frac{\ell_\mathcal{I}}{\ell^\Sigma_1} \triangleq n_\mathcal{I}$ channel uses to transmit $\ell_\mathcal{I}$ finite field symbols using random linear block codes. 

Now, since $R^\Sigma_\text{max} = R^\Sigma_1 =  \frac{\ell^\Sigma_1 \log_2 F}{n} = \frac{\ell_\mathcal{I} \log_2 F}{n_\mathcal{I}}$ satisfies \eqref{eq:common-capacity-1}, using Lemma~\ref{lemma:uplink} and choosing a sufficiently large $n$, the relay can reliably decode $f_\mathcal{I}$ in each block $\mathcal{I}$ and hence $\boldsymbol{f}$.

\subsection*{2) Each user is able to decode $\boldsymbol{f}$ broadcast by the relay in $n$ downlink uses.}

Since $\boldsymbol{f}$ is a finite-field vector of length $\ell^\Sigma_1$, there are at most $F^{\ell^\Sigma_1}$ distinct vectors $\boldsymbol{f}$. The relay selects some $p(x_0)$, randomly generates a single length-$n$ sequence $\boldsymbol{x_0}$ for each $\boldsymbol{f}$ according to $\prod_{t=1}^np(x_0[t])$, and transmits $\boldsymbol{x}_0(\boldsymbol{f})$ on the downlink.


Each user $a \in \{1,2,\dotsc, L\}$ attempts to decode $\boldsymbol{f}$ on the downlink with the help of its prior messages, $\{W_\mathcal{I}: a \in \mathcal{I}\}$. Recall that $\boldsymbol{f}$ is a deterministic function of all users' messages. Of all the messages, only $\{W_\mathcal{I}: a \notin \mathcal{I}\}$ are unknown to $a$. This means user $a$ needs to decode the actual transmitted $\boldsymbol{f}$ out of at most $F^{\ell^\Sigma_a} = 2^{nR^\Sigma_a}$ candidates. 
Denote this candidate set by $\mathcal{D}_a$, i.e., the set of distinct $\boldsymbol{f}$ that is formed by all possible $\{W_\mathcal{I}: a \notin \mathcal{I}\}$ and the correct $\{W_\mathcal{I}: a \in \mathcal{I}\}$. It contains all possible $\boldsymbol{f}$ user $a$ may decode to. We have
\begin{equation}
|\mathcal{D}_a| \leq |\{W_\mathcal{I}: a \notin \mathcal{I}\}| = F^{\ell^\Sigma_a}. \label{eq:size}
\end{equation}

Each user $a \in \{1,2,\dotsc, L\}$ decodes $\boldsymbol{f}$ from its received downlink channel outputs $\boldsymbol{Y}_a$ if it can find a unique vector $\boldsymbol{f} \in \mathcal{D}_a$ such that
\begin{equation}
(\boldsymbol{X}_0(\boldsymbol{f}), \boldsymbol{Y}_a) \in A^{(n)}_\eta(X_0,Y_a), \label{eq:typical}
\end{equation}
where $A^{(n)}_\eta(X_0,Y_a)$ is the set of jointly typical sequences $\{(\boldsymbol{x}_0,\boldsymbol{y}_a)\}$~\cite[p.\ 195]{coverthomas06}. Otherwise, user $a$ declares a decoding error. So, user $a$ makes an error in decoding if the event $E_1 \cup E_2$ occurs, where
\begin{itemize}
\item $E_1$: the correct $\boldsymbol{f} \in \mathcal{D}_a$ does not satisfy \eqref{eq:typical},
\item $E_2$: some wrong $\boldsymbol{f}'  \in \mathcal{D}_a$ satisfies \eqref{eq:typical}.
\end{itemize}

By definition, we have $\boldsymbol{f} \in \mathcal{D}_a$.
From the joint asymptotic equipartition property (JAEP)~\cite[Thm.\ 7.6.1]{coverthomas06}, we know that
$\Pr\{ E_1 \} \rightarrow 0$ as $n \rightarrow \infty$.

We now evaluate the probability of $E_2$:
\begin{subequations}
\begin{align}
\Pr \{E_2\} &= \Pr \{ \text{some } \boldsymbol{f}'  \in \mathcal{D}_a \text{ satisfies } \eqref{eq:typical} \} \\
&\leq \sum_{\boldsymbol{f}' \in \mathcal{D}_a \setminus \{\boldsymbol{f}\}} \Pr \{ \boldsymbol{f}' \text{ satisfies } \eqref{eq:typical} \} \\
& \leq  ( F^{\ell^\Sigma_a} -1) 2^{-n( I(X_0;Y_a) - 3 \eta)} \label{eq:decoding-1}\\
&< 2^{n( \frac{\ell^\Sigma_a \log_2 F}{n} - I(X_0;Y_a) + 3\eta) } \\
&= 2^{n( [R^\Sigma_a - I(X_0;Y_a) + 3\eta) }
\end{align}
\end{subequations}
where \eqref{eq:decoding-1} follows from \eqref{eq:size} and the JAEP~\cite[Thm.\ 7.6.1]{coverthomas06}, and $\eta >0$ defined in \eqref{eq:typical} can be chosen as small as desired.

As \eqref{eq:common-capacity-2} holds, $\Pr\{E_1 \cup E_2\} \leq \Pr\{E_1\} + \Pr\{E_2\} \rightarrow 0$ as $n \rightarrow 0$ by choosing a sufficiently small $\eta$. Thus each user $a$ can reliably decode $\boldsymbol{f}$. 

This technique of broadcast with receiver side information---where the receiver does not need to search over all possible $\boldsymbol{f}$ to find the correct one---has been used for the downlink of the two-way relay channel~\cite{oechteringschnurr08}, where there are two users.


\section{Conclusions}
In this paper we have derived optimal coding functions for the multi-way relay channel where the uplink is restricted to a finite-field with additive noise while the downlink is given by an arbitrary channel model and where users may have portions of their messages in common. The assumption of a finite-field uplink channel allows each user to transmit at the point-to-point capacity for the user-to-relay channel, allowing us to focus on designing the coding function that minimizes the information the relay needs to decode whilst still allowing the users to decode all required messages from the downlink broadcast.

\appendices

\section{Previous Coding Scheme for Only Private Messages is Suboptimal for Shared Messages} \label{section:example}



For the finite-field multi-way relay channel with only private messages (i.e., $R_{i,j} = 0$ for all $(i,j)$), we have the following capacity result, which is a straightforward extension of \cite{ongmjohnsonit11}:
\begin{lemma}\label{lemma:private}
Consider the finite-field multi-way relay channel with only private messages, where each user $i$ transmits an independent private message $W_i$. The rate tuple $(r_1,r_2,\dotsc,r_L) \triangleq \mathbf{r}_\textnormal{private-only}$ is achievable if there exists some $p(x_0)$ such that the following is true for all $i \in \{1,2,\dotsc, L\}$:
\begin{equation}
\sum_{j \in \{1,2,\dotsc, L\} \setminus \{i\}} r_j < \min \{ \log_2 F - H(N_0), I(X_0;Y_i) \}. \label{eq:capacity-private-messages}
\end{equation}
Conversely, if $\mathbf{r}_\textnormal{private-only}$ is achievable, then there exists some $p(x_0)$ such that  \eqref{eq:capacity-private-messages} is satisfied all $i \in \{1,2, \dotsc, L\}$ with a non-strict inequality.
\end{lemma}

In this paper, we refer to the coding scheme that achieves the capacity region in Lemma~\ref{lemma:private} as functional-decode-forward for private messages (FDF-P). 
Although we can use the FDF-P scheme when there are shared messages, the scheme is not always optimal. Consider three users, $L=3$, with shared messages, where
\begin{enumerate}
\item $X_1,X_2,X_3,Y_0,N_0 \in \{0,1,2,3\}$, i.e., $F=4$, 
\item $\Pr\{N_0=0\}=\Pr\{N_0=1\} = 1/2$,\\ $\Pr\{N_0=2\} = \Pr\{N_0=3\} = 0$,
\item $X_0, Y_1, Y_2, Y_3 \in \{0,1\}$,
\item $Y_1 = Y_2 = Y_3 = X_0$.
\end{enumerate}

Suppose that we have private messages $W_i$ (of rate $R_i$), each known to one user $i$, and shared messages $W_{i,j}$ (of rate $R_{i,j}$), each known to two users $i$ and $j$.
To transmit private messages and shared messages using FDF-P, we transmit them as if they are all private messages. To this end, we first split each shared message $W_{i,j}$ into two parts, say $W_i'$ and $W_j'$, and let user $i$ transmit $W_i'$ and user $j$ transmits $W_j'$. More specifically, we split the message $W_{1,2}$ into independent sub-messages, i.e., $W_1'$ of rate $R_1'$ and $W_2'$ of rate $R_2'$, where $R_1' + R_2' = R_{1,2}$. Similarly, we split the shared messages (i) $W_{1,3}$ into $W_1''$ and $W_3''$ with rates $R_1''$ and $R_3''$ respectively, and (ii) $W_{2,3}$ into $W_2''$ and $W_3'$ with rates $R_2''$ and $R_3'$ respectively. Doing this, each user $i$ will need to transmit $(W_i, W_i', W_i'')$ to the other two users at the rate $r_i = R_i + R_i' + R_i''$, where $R_i', R_i'' \geq 0$.

Suppose that the following rates are achievable by FDF-P: $R_1 = R_2 = R_3 = 0.4 - \delta$, $R_{1,2}=0.2-\delta$, $R_{1,3} = R_{2,3} = 0.15-\delta$ for some small $\delta>0$. 
From Lemma~\ref{lemma:private}, we have
\begin{align}
&\quad\; \; r_1 + r_2 = R_1 + R_2 + R_{1,2} + R_1'' + R_2'' \nonumber\\
&\quad\quad\quad\quad\;\; \leq \log_2 F - H(N_0) = 1,\\
&\Rightarrow (R_{1,3} - R_3'') + (R_{2,3} - R_3') = R_1'' + R_2''\nonumber \\
&\quad\quad \leq 1 - R_1 + R_2 + R_{1,2} = 3 \delta,\\
&\Rightarrow R_3' + R_3'' \geq R_{1,3} + R_{2,3} - 3\delta =   0.3 - 5\delta.
\end{align}
So, $r_1 + r_3 = R_1 + R_3 + R_3' + R_3'' + R_1' + R_1'' \geq 1.1 - 7\delta + R_1' + R_1'' \geq 1.1 - 7\delta$. 
Choosing $\delta = 0.01$, we have $r_1 + r_3 \geq 1.093$. But from Lemma~\ref{lemma:private}, we must have  $r_1 + r_3 \leq 1$ (contradiction). Hence, these rates are not achievable using the FDF-P, but one can show that they are achievable using the scheme proposed in this paper (see Theorem~\ref{theorem:main}).


\begin{thebibliography}{10}
\providecommand{\url}[1]{#1}
\csname url@samestyle\endcsname
\providecommand{\newblock}{\relax}
\providecommand{\bibinfo}[2]{#2}
\providecommand{\BIBentrySTDinterwordspacing}{\spaceskip=0pt\relax}
\providecommand{\BIBentryALTinterwordstretchfactor}{4}
\providecommand{\BIBentryALTinterwordspacing}{\spaceskip=\fontdimen2\font plus
\BIBentryALTinterwordstretchfactor\fontdimen3\font minus
  \fontdimen4\font\relax}
\providecommand{\BIBforeignlanguage}[2]{{%
\expandafter\ifx\csname l@#1\endcsname\relax
\typeout{** WARNING: IEEEtran.bst: No hyphenation pattern has been}%
\typeout{** loaded for the language `#1'. Using the pattern for}%
\typeout{** the default language instead.}%
\else
\language=\csname l@#1\endcsname
\fi
#2}}
\providecommand{\BIBdecl}{\relax}
\BIBdecl

\bibitem{ongmjohnsonit11}
L.~Ong, S.~J. Johnson, and C.~M. Kellett, ``The capacity region of multiway
  relay channels over finite fields with full data exchange,'' \emph{IEEE
  Trans. Inf. Theory}, vol.~57, no.~5, pp. 3016--3031, May 2011.

\bibitem{onglechnerjohnsonkellett13}
L.~Ong, G.~Lechner, S.~J. Johnson, and C.~M. Kellett, ``The three-user
  finite-field multi-way relay channel with correlated sources,'' \emph{IEEE
  Trans. Commun.}, vol.~61, no.~8, pp. 3125--3135, Aug. 2013.

\bibitem{elgamalkim2001}
A.~{El Gamal} and Y.~Kim, \emph{Network Information Theory}, 1st~ed.\hskip 1em
  plus 0.5em minus 0.4em\relax Cambridge University Press, 2011.

\bibitem{kramergastpar04}
G.~Kramer, M.~Gastpar, and P.~Gupta, ``Cooperative strategies and capacity
  theorems for relay networks,'' \emph{IEEE Trans. Inf. Theory}, vol.~51,
  no.~9, pp. 3037--3063, Sept. 2005.

\bibitem{nazergastpar11}
B.~Nazer and M.~Gastpar, ``Compute-and-forward: Harnessing interference through
  structured codes,'' \emph{IEEE Trans. Inf. Theory}, vol.~57, no.~10, pp.
  6463--6486, Oct. 2011.

\bibitem{gunduzgoldsmithpoor-13-it}
D.~G{\"u}nd{\"u}z, A.~Yener, A.~Goldsmith, and H.~V. Poor, ``The multiway relay
  channel,'' \emph{IEEE Trans. Inf. Theory}, vol.~59, no.~1, pp. 51--63, Jan.
  2013.

\bibitem{ahlswedecai00}
R.~Ahlswede, N.~Cai, S.~R. Li, and R.~W. Yeung, ``Network information flow,''
  \emph{IEEE Trans. Inf. Theory}, vol.~46, no.~4, pp. 1204--1216, July 2000.

\bibitem{namchunglee09}
W.~Nam, S.~Chung, and Y.~H. Lee, ``Capacity of the {G}aussian two-way relay
  channel to within $\frac{1}{2}$ bit,'' \emph{IEEE Trans. Inf. Theory},
  vol.~56, no.~11, pp. 5488--5494, Nov. 2010.

\bibitem{ongkellettjohnson12it}
L.~Ong, C.~M. Kellett, and S.~J. Johnson, ``On the equal-rate capacity of the
  {AWGN} multiway relay channel,'' \emph{IEEE Trans. Inf. Theory}, vol.~58,
  no.~9, pp. 5761--5769, Sept. 2012.

\bibitem{coverthomas06}
T.~M. Cover and J.~A. Thomas, \emph{Elements of Information Theory},
  2nd~ed.\hskip 1em plus 0.5em minus 0.4em\relax Wiley-Interscience, 2006.

\bibitem{mohajertiandiggavi10}
S.~Mohajer, C.~Tian, and S.~N. Diggavi, ``On source transmission over
  deterministic relay networks,'' in \emph{Proc. IEEE Inf. Theory Workshop
  (ITW)}, Cairo, Egypt, Jan. 6--8 2010.

\bibitem{covergamal79}
T.~M. Cover and A.~A. {El Gamal}, ``Capacity theorems for the relay channel,''
  \emph{IEEE Trans. Inf. Theory}, vol. IT-25, no.~5, pp. 572--584, Sept. 1979.

\bibitem{oechteringschnurr08}
T.~J. Oechtering, C.~Schnurr, and H.~Boche, ``Broadcast capacity region of
  two-phase bidirectional relaying,'' \emph{IEEE Trans. Inf. Theory}, vol.~54,
  no.~1, pp. 454--458, Jan. 2008.

\end{thebibliography}


\end{document}